\documentclass[12pt]{iopart}

\newcommand{\be}{\begin{equation}}
\newcommand{\ee}{\end{equation}}
\def\ba{\begin{aligned}}
\def\ea{\end{aligned}}

\newcommand{\bea}{\begin{eqnarray}}
\newcommand{\eea}{\end{eqnarray}}

\usepackage{color}
\usepackage{psfrag}
\usepackage[pdftex]{graphicx}
\usepackage{epsfig}

\usepackage[fit]{truncate}
 \usepackage{fancyhdr}
\begin{document}

\title{Multifractality and quantum-to-classical crossover in the Coulomb anomaly at the Mott-Anderson metal-insulator transition}

\author{M. Amini $^1$, V. E. Kravtsov $^{2,3}$  and M. M\"{u}ller $^2$}

\address{$^1$ Isfahan University of Technology, 84154-83111
Isfahan, Iran.\\$^2$ Abdus Salam International Center
for Theoretical Physics, P.O. Box 586, 34100 Trieste, Italy.\\
$^3$ L. D. Landau Institute for Theoretical Physics,
Chernogolovka, Russia}

\begin{abstract} 
 We study the interaction driven localization transition, which a
recent experiment in Ga$_{1-x}$Mn$_x$As has shown to come along with
multifractal behavior of the local density of states (LDoS) and the
intriguing persistence of critical correlations close to the Fermi
level. We show that the bulk of these phenomena can be understood
within a Hartree-Fock treatment of disordered, Coulomb-interacting
spinless fermions. A scaling analysis of the LDoS  correlation
demonstrates multifractality with correlation dimension
$d_{2}\approx 1.57$, which is significantly larger than at a non-interacting Anderson transition. At the interaction-driven transition the states
at the Fermi level become critical, while
 the bulk of the spectrum
remains delocalized up to substantially stronger interactions. The
mobility edge stays close to the Fermi energy in a wide range of
disorder strength, as the interaction strength is further increased.
The
localization transition is concomitant with the quantum-to-classical
crossover in the shape of the
 pseudo-gap in the tunneling density of states, and with the proliferation
 of metastable HF solutions that suggest the onset of a glassy regime with poor screening properties.



\end{abstract}








\pacs{71.30.+h PACS2: 64.60.al PACS3: 71.23.An}

\maketitle
\section{Introduction.}
The
recent tunneling experiments on Ga$_{1-x}$Mn$_x$As by
Richardella et al.~\cite{Yazdani} have demonstrated critical,
multifractal correlations in the local
tunneling density of states at the
metal-insulator transition in this semiconductor. While this would
be expected from the theory of Anderson localization in
non-interacting fermions, the experiment bears clear signs of the
relevance of electron-electron interactions. Most strikingly, the
critical correlations were found to persist very close to the Fermi
level, even upon doping further into the metallic regime. This
phenomenon clearly originates in interactions, which single out the
Fermi level as a special energy.

The effect  of interaction 
has been studied on either side of the metal-insulator (MI)
transition~\cite{50Anderson}, but rather little is known about its
role close to criticality. In a seminal work Efros and Shklovskii
 showed \cite{ES, ES-Springer}, that deep in the insulator the
Coulomb interactions between localized electrons create a pseudogap
in the density of states (DoS) near the Fermi level
$\varepsilon_{F}$, where the DoS vanishes as $\rho(\varepsilon)\sim
|\varepsilon-\varepsilon_{F}|^{2}$ in 3D. This is reflected in the
Efros-Shklovskii law of variable-range hopping conductivity at low temperatures,
$\ln \sigma(T)\propto -\sqrt{T_0/T}$. In the opposite limit of
weakly disordered metals, Altshuler and Aronov discovered
\cite{AA-Elsevier-book}
interaction corrections to both the DoS near $\varepsilon_F$ and the
low $T$ conductance. In particular, for spinless fermions, disorder
and repulsive interactions both enhance the tendency to localize. In
the weakly localized regime the tunneling DoS has a dip at
$\varepsilon_F$ with~\cite{AA-Elsevier-book}
$\delta\rho=\rho(\varepsilon_{F}+\omega)-\rho(\varepsilon_{F})\propto
\sqrt{\omega}$. However, in contrast to the classical Efros-Shklovskii pseudo-gap,
the Altshuler-Aronov corrections are of purely quantum (exchange) origin:
$\delta\rho\propto \hbar^{3/2}$, if the diffusion coefficient and
the Fermi-velocity are held fixed.

The quantum corrections of Ref.~\cite{AA-Elsevier-book} can be
effectively summed to obtain a non-perturbative result near $d=2$ by
using the formalism of the non-linear sigma-model due to
Finkel'stein \cite{Fin-review, Bel-Kirk}. An effective action
approach was suggested by Levitov and Shytov~\cite{LevShyt} and
Kamenev and Andreev \cite{KamenevAndreev} to derive the
non-perturbative expression for the tunneling DoS in a weakly
disordered 2D system near the Fermi energy. Remarkably, in the lower
critical dimension $d=2$, the DoS at $\varepsilon_F$ vanishes
exactly in the thermodynamic limit. In higher dimensions $d>2$
instead, the above results suggest the following qualitative picture
\cite{LMNS, Bel-Kirk, Vojta}: The pseudo-gap in the one-particle DoS
gradually grows with increasing disorder or repulsion strength.
$\rho(\varepsilon_F)$ eventually vanishes at the localization
transition, and remains zero in the insulator. The shape of the
pseudo-gap evolves from the quantum behavior $\delta\rho\propto
\sqrt{\omega}$  in the metal to some non-trivial power
$\rho(\varepsilon_{F}+\omega)\propto \omega^{\mu}$, $\mu>0$ at the
Anderson transition point, to the classical $\omega^{2}$ behavior in
the deep insulator. A power law suppressed density of states at criticality, $\rho(\omega)\propto \omega^\mu$ is also predicted within an $\epsilon$-expansion in $d=2+\epsilon$ dimensions \cite{Bel-Kirk}. However, the actual scenario might be more
complex if the localization transition is accompanied, or even
preceded by a transition to glassy or other density-modulated
phases, aspects, which so far have not been taken into account
by existing theories.

A fascinating property of electronic eigenfunctions near the
localization transition of non-interacting particles is their
multifractality~\cite{Mirlin-rep}. It is an exact property of
critical states at the mobility edge $\varepsilon=\varepsilon_m$,
but, also off-critical states exhibit multifractal character inside
their localization or correlation radius $\xi$~\cite{KrCue}. This
fractality has important effects on interactions (both repulsive and
attractive), as it enhances their local matrix elements. In the case
of predominantly attractive interactions, it may induce local
pairing gaps in weak Anderson insulators. In more conducting
systems, it may lead to enhanced superconducting transition
temperatures.~\cite{IFK, IFKCue, Bur, KrProc}.

It has remained an unresolved theoretical question whether such
subtle wavefunction correlations survive in the presence of Coulomb
interactions.
The reason to doubt their survival is most easily seen on the level of a Hartree-Fock (HF) approach,
where the combinations optimizing one-particle HF orbitals in the presence of interaction
are a linear combination of non-interacting wave functions of different energies.
If one na\"{\i}vely assumes that the fractal patterns of such wave functions are only weakly correlated,
one may expect partial or even complete degradation of the fractal structure in the HF wavefunctions, due to a superposition of a large number of random uncorrelated fractal patterns. In reality the fractal patterns of non-interacting wave functions are strongly correlated even at large energy separation ~\cite{IFK, IFKCue, KrProc}. Nevertheless, the question remains whether such interaction-induced superpositions give rise to a change of the fractal dimension of HF wave functions or may destroy fractality completely, despite of the correlations. There is indeed a subtle interplay between the strength of correlations and the effective number of non-interacting wave functions which superpose in the HF wavefunctions.
Another na\"{\i}ve argument, advocating the opposite conclusion, puts forward that the HF Hamiltonian is essentially a one-particle Hamiltonian of the same basic symmetry as for the non-interacting case. Hence, invoking universality, one would expect the same statistics of both non-interacting and HF wave functions. The flaw in this argument is that the matrix elements of the HF Hamiltonian which are {\em self-consistently} determined, possess correlations which could be long-range in the presence of badly screened long-range interactions between particles.   Thus the two different na\"{\i}ve arguments lead to two opposite conclusions. According to one of them the fractal pattern of the HF wave function should be smeared out while the other one advocates unchanged fractal patterns. One of the main results of this study is to show that the first argument is in fact closer to reality.

On the other hand, the direct observation of
multifractality in the tunneling spectra of Ref.~\cite{Yazdani}
strongly suggests that the mulfifractality survives interaction.
Indeed, the measured auto-correlation function
of the local DoS (LDoS) showed a well-established power-law decay
with distance on the sample surface. Surprisingly, this critical
behavior appeared to be nearly pinned to the Fermi energy without any
fine-tuning of the ${\rm Mn}$ impurity concentration, implying that
the mobility edge $\varepsilon_m$ remains close to $\varepsilon_F$
in a broad range of disorder strengths. As mentioned above this
indicates the importance of interactions, since they single out
$\varepsilon_{F}$ as the center of the pseudogap. In this Letter, we
address the problem of multifractality at the interacting
localization transition theoretically, and study the mechanism by
which interactions pin the mobility edge $\varepsilon_m$ nearly  to
$\varepsilon_F$ in a broad parameter range.

As a na\"{\i}ve rationale for this pinning of the mobility edge one
may consider that localization occurs earlier
where the density of states is lower. Thus localization is naturally
prone to occur first within the interaction-induced pseudogap,
making $\varepsilon_m$ track $\varepsilon_F$ rather closely.
However, it is only the single-particle (tunneling) DoS that has a
pronounced dip near $\varepsilon_{F}$, while the global
thermodynamic DoS, $dn/d\mu$, that enters the conductivity via the
Einstein relation, usually shows a different behavior. Hence, it is
not obvious which notion of DoS is relevant for localization and
transport purposes (cf. discussions in \cite{BES84,LMNS,Lee} about
global vs. local $(dn/d\mu)^{-1}$). A more detailed analysis is thus
required in order to show that indeed the LDoS close to
$\varepsilon_F$ can be critical, while in the bulk of the spectrum
the correlations are still metallic.

\subsection{Model}
To address these questions we consider a model of spinless fermions on a 3D
cubic lattice of size $N=10^3$ with the tight-binding Hamiltonian
\be \label{tbH}
H_{0}=\sum_{i}(\epsilon_{i}-\mu)\,c^{\dagger}_{i}c_{i}-t
\sum_{\langle ij \rangle}c^{\dagger}_{i}c_{j}+{\rm h.c.}, \ee
interacting via long-range Coulomb repulsion:
\be\label{H-int}
H_{1}=\frac{U}{2}\sum_{i,j}\frac{n_{i}n_{j}}{r_{ij}}. \ee
We employ periodic boundary conditions and choose $t=1$ as the unit
of energy. The on-site energies $\epsilon_{i}$ are random,
independently and uniformly distributed in $\epsilon_{i}\in
[-\frac{W}{2},\frac{W}{2}]$. The chemical potential $\mu$ depends on
interaction and is chosen so as to keep the average density $1/2$.
For non-interacting particles ($U=0$) the localization transition is
known to occur at the disorder strength $W_{c}=16.5$ in this model
\cite{16-5}. In the present work we choose $W=14 < W_{c}$ (not
particularly close to $W_{c}$), so as to mimic conditions of
Ref.~\cite{Yazdani} where the impurity concentration was not
specially tuned.

We attack this problem numerically by considering the interactions
in the Hartree-Fock (HF) approximation. This amounts to studying an
effective single-particle model with self-consistent on-site
energies and hopping amplitudes. In order to clarify the role of long-range interactions, we truncated the Coulomb interaction at a finite range, and then  progressively increased its range up to the
size $L=10$ of the 3D system, defining $r_{ij}$ as the shortest distance on the torus.
We first took into account the Hartree
terms (occupation numbers $n_{j}$ in the sum
$U\sum_{j}r_{ij}^{-1}\,n_{j}$) and the Fock terms (expectation values of
$c^{\dagger}_{i}c_{j}$)  up to the $5^{\rm th}$ nearest
neighbors. Then we considered the Fock terms up to the $5^{\rm th}$ nearest
neighbors while the Hartree terms were considered up to the $20^{\rm th}$ nearest
neighbors. Finally we  tackled the full self-consistent problem for all neighbors.

\subsection{Overview of results}
The main result of our paper is to establish the
persistence of multifractality in the presence of full-range Coulomb interaction.
Notably, the fractal
dimension we find, $d_{2}\approx 1.57\pm 0.05$, appears to be significantly larger than in the non-interacting case.
With decreasing range of interaction the effective $d_{2}$ in a finite sample decreases ($d_{2}=1.38\pm 0.05$ for interaction up to the $5^{\rm th}$ nearest neighbor) until it reaches its value $d_{2}=1.35\pm 0.05$
for the non-interacting case. This marks  essential
progress in comparison to earlier works based on the HF approach~\cite{Vojta}. As we will
describe below, the critical behavior exhibits various further
interesting features that are specific to the interacting case. Most
importantly, we establish that within the insulating
phase, even considerably far from the metal-insulator transition, the mobility edge remains very close to the Fermi level.

Further, we study the evolution of the pseudo-gap in the HF density of states (DoS) $\rho(\omega)$ as the increasing interaction drives the system towards the localization transition. In particular we confirm (within our accuracy) the scaling relationships suggested by McMillan and Shklovskii ~\cite{McMillan,LMNS} which relate the critical power law of the pseudo-gap $\rho\propto \omega^\mu$ with the dynamical scaling exponent $\eta$ and the exponent that describes the dependence of the static dielectric constant  $\kappa_{0}\propto \xi^{\eta-1}$ on the localization radius $\xi$ in the insulator phase.

Finally, for the first time we address the question of multiplicity of HF solutions, and the competition of the related glassy features and localization. We show that within our accuracy the onset of multiplicity of solutions with increasing interaction strength (an indication of an emerging glassy energy landscape) coincides  with the localization transition. In contrast, the charge ordering (typical for a Mott transition) occurs at much stronger interaction. Thus we argue that the localization transition should better be called  an Anderson-glass transition, rather than an Anderson-Mott transition.

A preliminary version of this paper was published as a preprint \cite{cond-mat}.

\section{Hartree-Fock calculations}
The effective Hartree-Fock (HF) Hamiltonian which corresponds to the
model given in Eqs.~(\ref{tbH},\ref{H-int}) is:
\be\label{HF-H}
H_{\rm HF}=\sum_{i}\tilde{V}_{i}\,c^{\dagger}_{i}c_{i}-\sum_{ij}\left( \tilde{t}_{ij}\,c^{\dagger}_{i}c_{j}+{\rm h.c.}\right)\,.
\ee
Here
\be\label{tildeV} \tilde{V}_{i}=\epsilon_{i}+\sum_{j}\frac{U}{|{\bf
r}_{i}-{\bf r}_{j}|}\,\langle c^{\dagger}_{j}c_{j}\rangle_{0}-\mu,
\ee
\be \label{tilde-t} \tilde{t}_{ij}=t_{ij}+ \frac{U}{|{\bf
r}_{i}-{\bf r}_{j}|}\,\langle c^{\dagger}_{j}c_{i}\rangle_{0}, \ee
where $t_{ij}=t$   is the bare nearest-neighbor hopping and $\langle
...\rangle_{0}$ denotes the quantum-mechanical ground-state
expectation value evaluated on the Slater determinant formed by the
lowest $N/2$ HF levels. The effective on-site energy $\tilde{V}_{j}$
contains the interaction-induced {\it Hartree} term which leads to
correlated on-site energies (potentially at long range), while the effective hopping
$\tilde{t}_{ij}$ contains the {\it Fock} term which may be
long-range as well.
We carried out calculations on a cubic 3D lattice of size $L=10$, using 3 ranges of interactions.
The first one took into account up to the 20$^{\rm th}$ nearest neighbors (460  sites $j$
nearest to $i$) in the Hartree term and Fock terms corresponding
to the 5$^{\rm th}$ nearest neighbors (the 56 nearest sites up to distance
$\sqrt{5}$).
A second calculation restricted the
Hartree and the Fock terms  equally to the 5$^{\rm th}$ nearest neighbors
in order to check the importance of Hartree terms and to ensure a correct implementation of the Pauli principle. A third calculation performed the self-consistent Hartree-Fock calculations with the full range of Coulomb interactions.

Note that the role of the Hartree and Fock terms is not the same in the metal and the insulator.
Deep in the insulator side, it is important to
keep the longer range Hartree terms to obtain the full classical
Efros-Shklovskii gap, while long range Fock terms are negligible due
to strong localization of the wavefunctions. In the metal, however, the role of
the Fock terms is expected to be more significant, while the Hartree terms incorporate an effective screening at long distances.

\subsection{Numerical implementation}
Even though completely standard, we briefly review the main steps involved in finding solutions of the HF equations.
The set $X$ of parameters  to be
found self-consistently comprises all the $\langle
c^{\dagger}_{i}c_{j}\rangle_{0}$, ($\sim L^{6}/2$ parameters for the
full-scale Coulomb interaction) plus the
$L^{3}$ diagonal parameters $\langle n_{i}\rangle_{0}=\langle
c^{\dagger}_{i}c_{i}\rangle_{0}$ (i.e. $\sim 500.000$ parameters for $L=10$).
The chemical potential $\mu$ is
always adjusted to assure half filling, as described below.

In order to find a self-consistent solution we begin with a
random initial guess for all the parameters $X_{\rm in}^{(0)}$
satisfying the condition
\be \label{fixN} \sum_{i}\langle n_{i}\rangle_{0}={\cal N}_{e}, \ee
where the number of particles ${\cal N}_{e}=N/2$ is fixed in our
calculation (half-filling). Diagonalizing the effective Hamiltonian
Eqs.~(\ref{HF-H})-(\ref{tilde-t}) using the initial $X_{\rm
in}^{(0)}$ one obtains the eigenfunctions $\psi_{m}({\bf r})$ and
eigenvalues $\varepsilon_{m}$, from which we compute  the {\it
output} parameters $X_{\rm out}^{(0)}$:
\bea\label{self-con} \langle
n_{i}\rangle_{0}&=&\sum_{m}|\psi_{m}({\bf
r}_{i})|^{2}\,f(\varepsilon_{m}), \\ \langle
c^{\dagger}_{i}c_{j}\rangle_{0}&=&\sum_{m}\psi_{m}^{*}({\bf
r}_{i})\psi_{m}({\bf r}_{j})\,f(\varepsilon_{m}), \eea
where $f(\varepsilon)$ is the Fermi distribution function with the
chemical potential $\mu$. It is to be found from the condition:
\be \label{chem} {\cal N}_{e}=\sum_{m}f(\varepsilon_{m}). \ee
At $T=0$ considered in this paper there is an uncertainty of the
position of $\mu$ between the two energy levels $\varepsilon_{{\cal
N}_{e}+1}$ and $\varepsilon_{{\cal N}_{e}}$. For most of the
calculations we have chosen $\mu = \frac{1}{2}(\varepsilon_{{\cal
N}_{e}+1}+\varepsilon_{{\cal N}_{e}})$. However, to avoid an artificial hard minigap at
 the bottom of the DoS dip, to study the latter we used a parametric mixing
$\mu=(1-a)\,\varepsilon_{{\cal N}_{e}+1}+a\,\varepsilon_{{\cal
N}_{e}}$ with $0<a<1$; $a$ was fixed for a given disorder
realization, but taken at random for different disorder
realizations.

An updated set of parameters to be used as initial parameters
$X_{\rm in}^{(n+1)}$ for the next, i.e., $(n+1)$-th iteration is chosen as
follows ($n=0,1...$):
\be\label{update}
X_{\rm in}^{(n+1)}=(1-\alpha)\,X_{\rm in}^{(n)}+\alpha\,X_{\rm out}^{(n)}.
\ee
The parameter $\alpha\in[0,1]$ is chosen such that the iteration
process is stable and leads to a convergent solution. The iteration
procedure is terminated and the output set of parameters is taken as
the converged solution if the absolute value of the difference
between the values of all the parameters $X$ of the previous and the
final iteration is less than $10^{-4}$ for the truncated Coulomb interaction, and
$10^{-5}$ for the full range Coulomb interaction. Once a converged solution is obtained, and the final set of
HF eigenfunctions $\psi_{m}({\bf r})$ and eigenvalues
$\varepsilon_{m}$ have been calculated, one can compute any quantity
expressible in terms of $\psi_{m}({\bf r})$ and $\varepsilon_{m}$.
The procedure is then repeated for different realizations of
disorder to obtain disorder averaged quantities, such as the DoS and the LDoS correlation functions.

We point out that the solution of the HF equations is unique only at small enough $U$ within the metallic regime. At large $U$, one expects a number of solutions that grows exponentially with the volume. In this regime, we analyze {\em typical} solutions of the HF equations, without optimizing the HF energy among different solutions. This choice will be discussed and justified in Sec.~\ref{s:glass}.

The typical number of iterations needed to obtain a HF solution for one realization of
disorder was $\sim 2000$ for the full-scale Coulomb interaction. The total computational time to obtain one HF solution was mostly limited by the time $\sim L^{9}$ of diagonalization of a matrix Hamiltonian
of the size $L^{3}\times L^{3}$ needed in each iteration. With a typical number of disorder realizations $\sim 2000$ the total time at $L=10$ was of the order of $1000/(\#cores)$ hours for each parameter set of interaction and disorder strengths. For all values ($\sim 20$) of interaction strengths $U$ necessary for our scaling analysis and the average number of cores $\sim 50$ used for parallel computing the total time was of the order of 400 hours for $L=10$.

\section{McMillan-Shklovskii scaling.}
The metal insulator transition in disordered systems is expected to occur as a second order phase transition at some interaction strength $U_c$. Close to criticality, where $\tau\equiv |1-U/U_c|\ll1$, one expects a scaling form for the density of states
as
\bea\label{f-rho} \rho(\omega\equiv
\varepsilon-\varepsilon_F)=\Delta^{-1}\; f_\rho(\omega/\delta), \eea
with
\bea\label{gamma-over-mu} \Delta\propto \tau^{-\gamma},
\;\;\;\;\delta \propto \tau^{\gamma/{\mu}}. \eea
In the critical regime,
\bea\label{large-x} f_\rho(|x|\gg 1)&\sim& |x|^{\mu}, \eea
whereas in the metal and the insulator,
\bea\label{small-x}
f_{\rho,M}(|x|\ll 1)&=&{\rm const.},\\
f_{\rho,I}(|x|\ll 1)&\sim& x^{2}, \eea
which capture the shape of the Altshuler-Aronov and Efros-Shklovskii pseudogaps as limiting cases.
The exponents in Eq.~(\ref{gamma-over-mu}) are chosen such that the dependence on the critical parameter $\tau$ disappears at criticality.

The scaling \cite{LMNS,McMillan} is based  on an assumption about the potential of a point
charge within the critical regime, i.e., at a distance $a\ll r\ll
\xi$. Here $\xi$ is the correlation length which diverges at the
transition as,
\be\label{xi-nuu}\xi\propto |1-U/U_{c}|^{-\nu}, \ee
and $a$ is a certain microscopic length (e.g., the distance between
donors in doped semiconductors \cite{LMNS}). In our simulations it
can be taken equal to the lattice spacing. The assumption is that
the potential behaves as a modified power law,
 \be\label{screening} V(r)\sim
U\, \left( \frac{a}{r}\right)^{\eta},\;\;\;\;U\sim e^{2}/a, \;\;\;\;\;a\ll r \ll \xi. \ee
As Eq.~(\ref{screening}) is essentially the relationship between the
length scale $r$ and the energy (or inverse-time) scale $V$, the exponent $\eta$
should coincide with the dynamical scaling exponent $z$. For the
non-interacting case the dynamic exponent takes its maximum value
$\eta = d=3$, while the minimal theoretically admissible value
cannot be smaller than the exponent of the Coulomb
potential~\cite{McMillan}, $\eta\geq 1$.

The exponent $\eta$ also governs the scaling of the static
dielectric constant in the insulator~\cite{McMillan}:
\be \label{Macmillan} \kappa_{0}\propto
(1-U/U_{c})^{-\zeta},\;\;\;\;\zeta=\nu\,(\eta-1). \ee
To show this it is enough to assume that in the insulator at
distances $r\gg\xi$ the potential $V(r)$ takes the usual form
of dielectric screening,
\be \label{diel-scr} V(r)=\frac{e^{2}}{\kappa_{0}\,r},\;\;\;\;r\gg
\xi \ee
and matches with the potential in Eq.~(\ref{screening}) at distances $r\sim \xi$.

The characteristic energy scale $\delta$ in Eq.~(\ref{f-rho}) is set by the
potential $V(r)$ at $r\sim \xi$:
\be\label{delta-eta} \delta=V(\xi) = e^{2}a^{\eta-1}\,\xi^{-\eta}.
\ee
In a system of finite size $L$, in the critical region
where $\xi\gg L$, one should replace $\delta$ by the
mean level spacing $\delta_{L}\sim V(L)$ .

Finally, the characteristic scale $\Delta$ of the DoS can be
expressed through $\delta$ and $\xi$ by a relationship  following from dimensional arguments:
\be \label{Delta-xi}
\Delta^{-1}=\frac{1}{\delta\,\xi^{3}}=\frac{\xi^{-(3-\eta)}}{U\,a^{\eta}},\;\;\;U\equiv
e^{2}/a. \ee
Eqs.~(\ref{delta-eta},\ref{Delta-xi}), as well as the scaling
Eq.~(\ref{f-rho}) are valid for:
\be \label{limits} a\ll \xi,\;\;\;\;\;U\gg \omega\gg
\delta_{L}\equiv U\, (a/L)^{\eta}. \ee
From Eqs.~(\ref{delta-eta},\ref{Delta-xi}) and (\ref{gamma-over-mu})
one immediately obtains the following scaling relations~\cite{McMillan,LMNS}:
\be \label{Shkl} \gamma=\nu(3-\eta),\;\;\;\;\mu=\frac{3}{\eta}-1,
\ee
in terms of $\eta$ and the correlation length exponent  $\nu$.

Note that the above scaling assumes only one critical scale $\xi$ separating different regimes of
$V(r)$. Should an additional scale (e.g.,  one related with a "screening transition") appear, the exponent $\mu$ will be independent of the dynamical scaling exponent $\eta$.

In the next section we analyze the evolution of the density of states across the transition in the light of the above scaling assumptions.

\section{Pseudo-gap in the density of states (DoS).}
\begin{figure}[h]
\center{\includegraphics[width=1\linewidth]{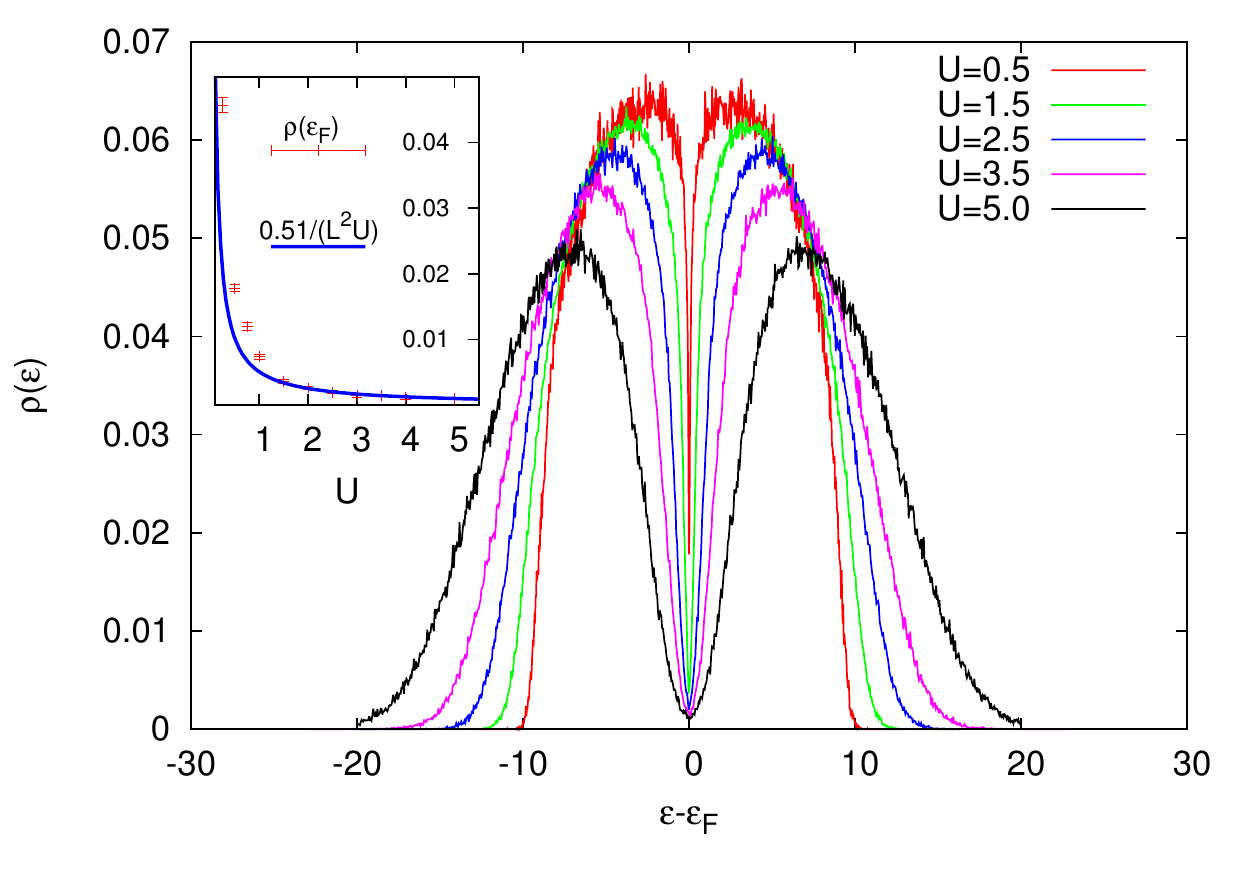}}
\label{fig:DoS}
\caption{(Color online)
Disorder-averaged DoS $\rho(\varepsilon)$ of
the HF states at $W=14$ at different interaction strengths $U$. The
crossover from the quantum Altshuler-Aronov correction $\delta\rho\sim
\sqrt{|\varepsilon-\varepsilon_{F}|}$ to the classical Efros-Shklovskii gap
$\rho\sim (\varepsilon-\varepsilon_{F})^{2}$ is seen. The bandwidth progressively increases with increasing $U$. - Insert:
$\rho(\varepsilon_{F})$ as a function of $U$. For $U>1.5$ the DoS
$\rho(\varepsilon_{F})\approx 0.5/(UL^{2})$ follows the classical Efros-Shklovskii
law, where $L=10$ is the system size.} \label{Fig:DoS}
\end{figure}
In Fig.~\ref{Fig:DoS} we present the DoS of
the HF levels.
One can see that deep in the metallic and in the insulating
regimes, HF correctly captures the Altshuler-Aronov and Efros-Shklovskii pseudogap features
discussed above, while it provides a non-trivial mean field approach
to describing various interesting phenomena happening at and close
to the MI transition.   The curvature of $\rho(\varepsilon)$ at small
$\omega=\varepsilon-\varepsilon_{F}$ is seen to change sign as $U$
increases.

From RG and scaling arguments~\cite{McMillan,LMNS} as presented above, one expects a critical power law
$\rho(\omega)\sim\omega^\mu$ in a frequency regime where $\omega>1/\rho(\omega)\xi^d$.
 \begin{figure}[h]
\center{
\includegraphics[width=0.7\linewidth]{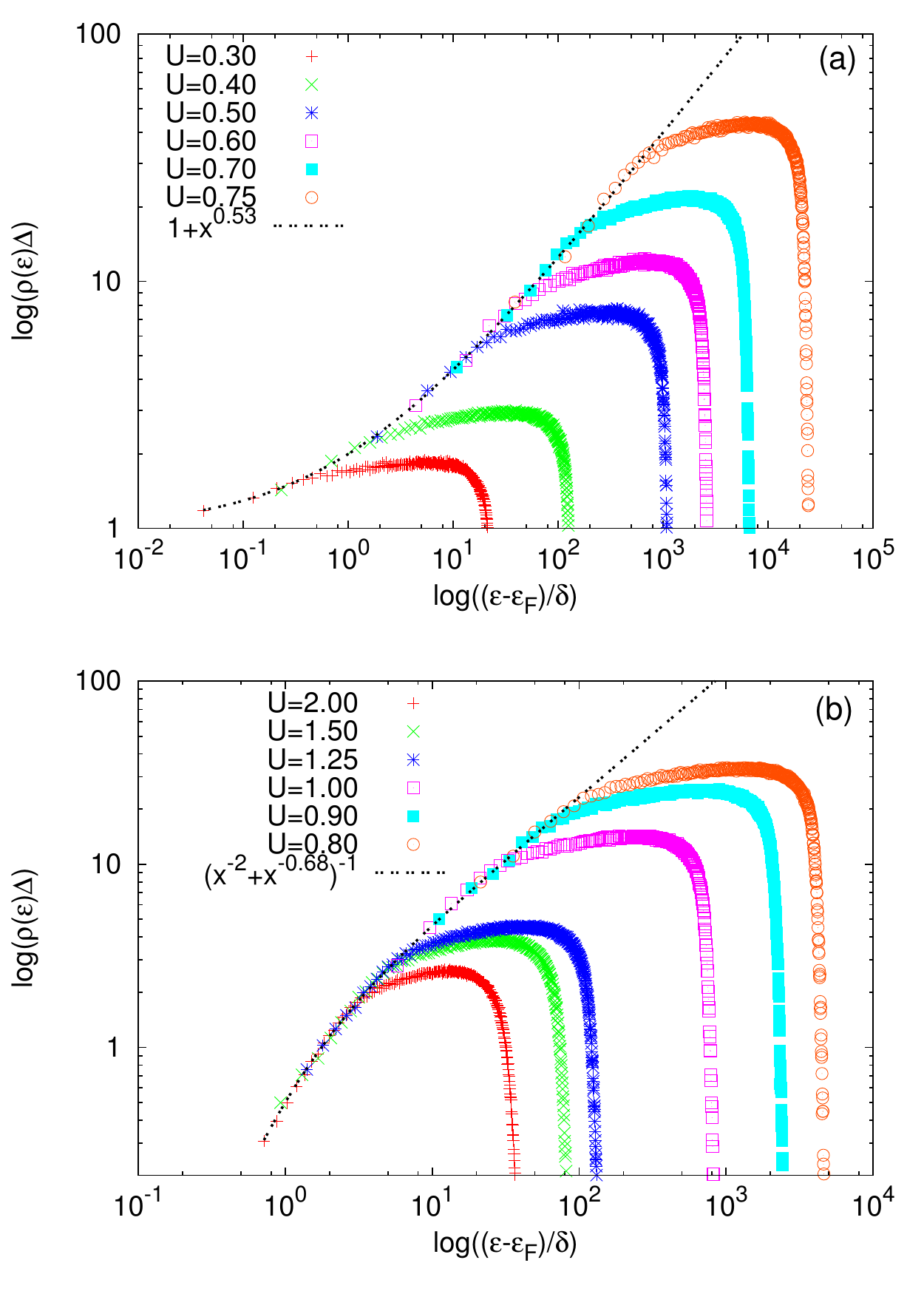}}
\caption{Collapse of data for the DoS in the window $U\gg
\omega\gg\delta_{L}$ (Eq.~(\ref{limits})) onto the scaling function
$f_{\rho}(x)$: (a) metallic side with $f_{\rho,M}(x)=1+x^{0.53}$ (b)
insulating side  with $f_{\rho,I}(x)=[x^{-2}+x^{-0.68}]^{-1}$. }
\label{fig:collapse}
\end{figure}

\begin{figure}[t]
\center{\includegraphics[width=0.5\linewidth]{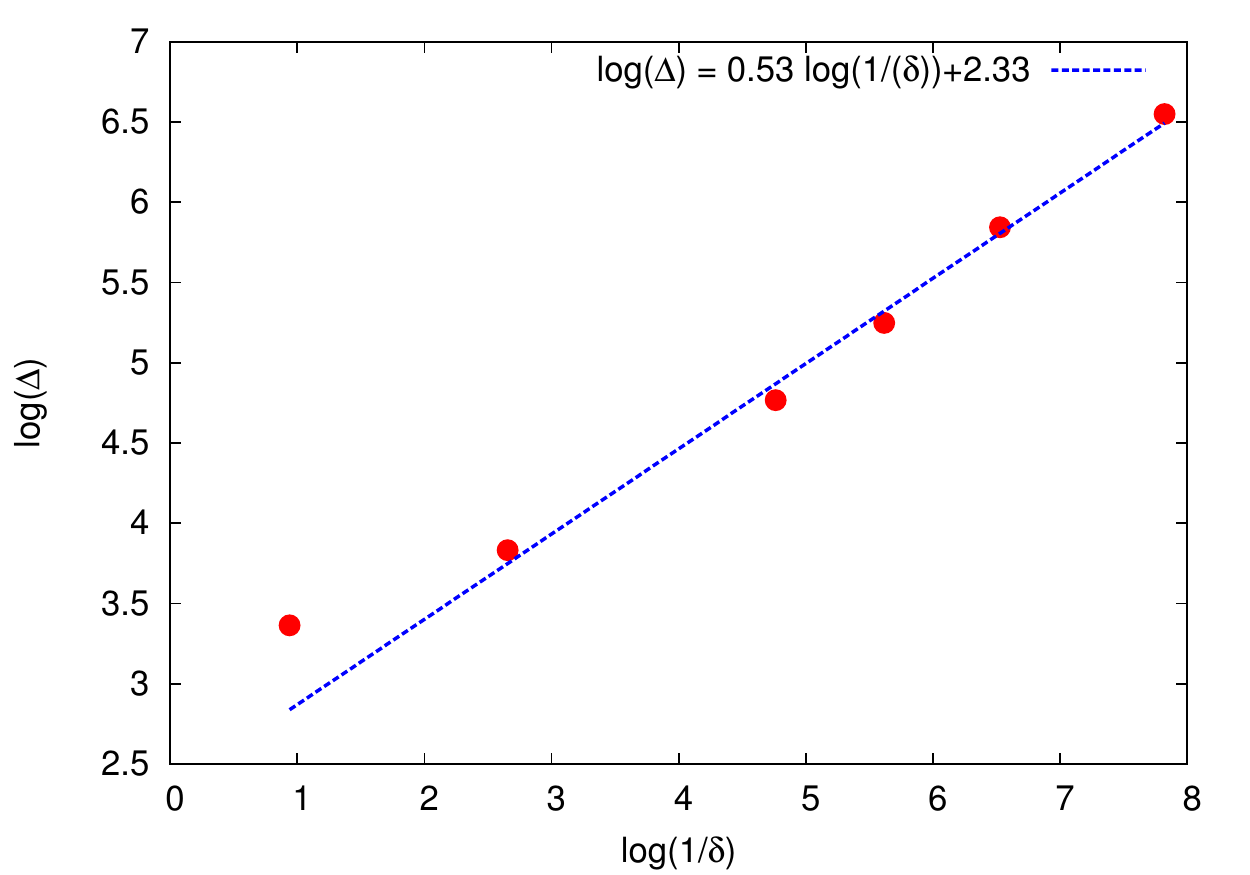}
\caption{Log-Log plot of $\Delta$ vs. $1/\delta$, as obtained from
the metallic side of the data collapse for the DoS. It is almost
linear with the slope $\mu_M=0.53$. }} \label{fig:loga-logb}
\end{figure}

In Fig.~\ref{fig:collapse} we verified the scalings
 Eqs.~(\ref{f-rho},\ref{large-x},\ref{small-x})
by collapsing the "low-energy" data (in the regime~(\ref{limits})) for
$\rho(\varepsilon)$ close to $\varepsilon_{F}$ onto the universal
scaling functions $f_{\rho,M}(x)$ and $f_{\rho,I}(x)$. From the
power-law behavior of $f_{\rho,M}$ at $x\gg 1$ we found for the
exponent $\mu$:
\be\label{mu-met} \mu_{M}=0.53, \ee
very close to the value experimentally observed in the tunneling DoS close to criticality~\cite{LeePRL}. We
cross-checked this result in Fig.~\ref{fig:loga-logb} by plotting
$\ln\Delta$ versus $\ln(1/\delta)$. We obtained an almost linear curve
in accordance with Eq.~(\ref{gamma-over-mu}), with the same slope
$\mu_M=0.53$ as in Fig.~\ref{fig:collapse}(a).
In the insulator the best collapse corresponds to:
\be\label{mu-ins} \mu_{I}=0.68, \ee
but the reliability of this exponent is not as high as the one on the
metallic side (for instance a test similar to Fig.~\ref{fig:loga-logb}
yields a slope 0.54, consistent rather with $\mu_{M}$, but smaller than
$\mu_{I}=0.68$). However, to be conservative we may conclude that
the critical exponent $\mu$ lies in the range:
\be\label{mu-mu} \mu=0.60\pm 0.15. \ee
From the obtained exponent $\mu$    we can estimate the dynamical
scaling exponent $\eta$:
\be\label{eta} \eta=\frac{3}{1+\mu}=1.9\pm 0.2. \ee
This yields the exponent $\zeta=0.9\pm 0.2$ characterizing the divergence of the dielectric constant in Eq.~(\ref{Macmillan}),
in a reasonably good agreement with the experimental value
\cite{LMNS} $\zeta\approx 0.71$.~\footnote{the exponent $\nu\approx 1.0$ in the insulator is obtained in the next section}

Thus we conclude that our results are compatible with the McMillan-Shklovskii scaling as well as with the available experimentally obtained values of the exponents $\mu$ and $\zeta$.

\section{ Auto-correlation of the local DoS and the fractal dimension $d_{2}$.}
\begin{figure}[h]
\center{\includegraphics[width=0.7\linewidth]{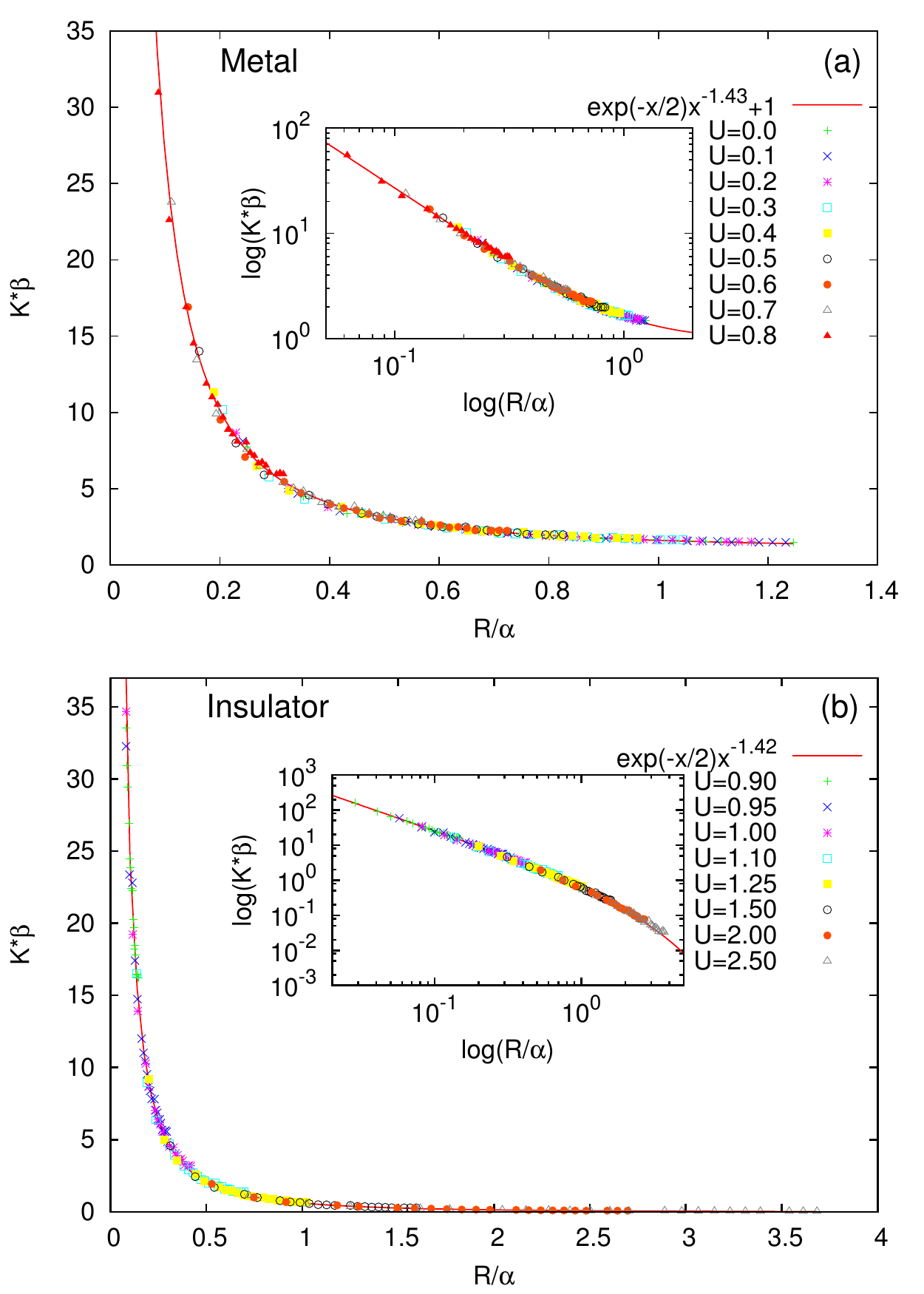}}
\caption{(Color online) Data collapse of the auto-correlation
of the LDoS, $K(R;\varepsilon_{F})$  at the Fermi energy for disorder strength
$W=14$ and a full-scale Coulomb interaction in a sample with $L=10$
onto (a) metallic  and (b) insulating  scaling functions $f_{M,I}$
[Eqs.~(\ref{f-met},\ref{f-ins})]. We find $d_2=1.57\pm 0.05$. The control calculations for non-interacting system with the same protocol and the same system size gave $d_{2}=1.34\pm 0.05$. }\label{fig:K-collapse}
\end{figure}
To study multifractality of the local DoS, we have computed the spatial correlations of
the HF wavefunctions $|\psi_{n}({\bf r})|^{2}$:
\be \label{K-r} K(R;\varepsilon)=\frac{\left\langle
 \sum_{n,\varepsilon_{n}\in\Omega(\varepsilon)}\sum_{{\bf r}}|\psi_{n}({\bf r})|^{2}|\psi_{n}({\bf r+R})|^{2}
 \right\rangle}{\left\langle\sum_{n,\varepsilon_{n}\in\Omega(\varepsilon)}\sum_{{\bf r}}|\psi_{n}({\bf r})|^{4}\right\rangle}, \ee
where $\varepsilon_{n}$
are the associated eigenvalues,
$\langle ...\rangle$ denotes the ensemble average over random
realizations of on-site energies $\epsilon_{n}$,  and $\Omega(\varepsilon)$ is a narrow interval of energies
of the order of the mean level spacing $\delta$, centered at
$\varepsilon$. Multifractal correlations \cite{IFKCue, Mirlin-rep} imply that in the range of distances
$\ell_{0}<R<\xi$ the correlation function is the same as at the
localization transition, $K(R;\varepsilon)\sim
(\ell_{0}/R)^{d-d_{2}}$. Here $\xi$ is the localization or correlation length
which diverges at the transition, and $\ell_{0}$ is
of the order of the lattice constant. $d=3$ is the dimensionality of space
and $d_{2}<d$ is the correlation fractal dimension. For $R>\xi$ the
correlation function $K$ distinguishes delocalized and localized regimes, saturating
to a constant in a metal and decreasing exponentially in an insulator. Close to
criticality one expects scaling behavior, that is: the correlations should collapse
to a single curve upon rescaling $R$ by $\alpha$ (a finite size corrected version of $\xi$),
and amplitudes by $\beta$, and expressing $K(R;\varepsilon)=\beta^{-1}\,f_{M,I}(R/\alpha)$,
where $f_{M,I}(x)$ are universal scaling functions on the metallic and insulating sides of the transition, respectively.
In order to optimize the choice of $\alpha, \beta$ (which both depend on $U$ and $\varepsilon$, while $W=14$ is fixed) and to
determine the correlation dimension $d_{2}$
 we use a simple analytical ansatz for $f_{M,I}$, which
captures the multifractal characteristics of the eigenfunction
correlations:
\bea \label{f-met} f_{M}(x) &=& x^{-(d-d_2)}\,e^{-x/B}+1, \\
\label{f-ins} f_{I}(x)&=& x^{-(d-d_2)}\,e^{-x/B}.  \eea
The fits were optimized by the value $B\approx 2$.

We first discuss the DoS correlations $K(R;\varepsilon_F)$ at the Fermi
level, upon varying the strength of the interaction $U$, cf.~Fig.~\ref{fig:K-collapse}.
The good quality of the collapse demonstrates that the behavior of
$K(R;\varepsilon)$ is consistent (using full-range Coulomb interactions) with multifractal correlations
with dimension:
\be
d_{2}\approx 1.57\pm 0.05, \;\;\;\;{\rm full-range}\;\;{\rm Coulomb}.
\ee
This is significantly larger than the fractal dimension found for the non-interacting case in the limit of large sample sizes
$d_{2}=1.29\pm 0.05$~\cite{Mirlin-rep} from the multifractal analysis of the moments of $|\psi_{n}({\bf r})|^{2}$. To gauge the finite-size effects in the non-interacting case we performed calculations of the correlation function $K(R;\varepsilon_{m})$ with collapse of the data for different disorder strengths $W$ (and $U=0$) similar to Fig.\ref{fig:K-collapse}. This yielded the effective $d_{2}=1.34\pm 0.05$ for the sample size $L=10$. From this we conclude that the presence of full-range Coulomb interactions strongly affects the multifractal correlations at the Fermi level, which are governed by a new interacting critical point with a correlation fractal dimension $d_{2}$ {\it larger} than for the non-interacting case.

This result is in full agreement with the qualitative picture outlined in the Introduction. It is also in line with recent results obtained via the $\epsilon=d-2$ expansion  in the unitary ensemble \cite{BurMirGor}. According to that study:
\be\label{eps-d-2}
d_{2}^{{\rm inter}}=2-\frac{\epsilon}{2},\;\;\;\;d_{2}^{{\rm non-inter}}=2-\sqrt{2\epsilon}.
\ee
Although the $\epsilon$ expansion fails to give an accurate prediction for $d=3$, the tendency for $\epsilon\leq 1$ is clearly that $d_{2}^{{\rm inter}}>d_{2}^{{\rm non-inter}}$.
An increase of $d_2$ is also expected from studies of systems with frustrating
interactions on the Bethe lattice, where the Efros-Shklovskii- (or Hartree-) type
suppression of the density of states around the chemical potential is found to
reduce the abundance of resonances (i.e., small denominators in the locator expansion). Therefore, the wavefunctions have less tendency to follow rare paths to increase the number of resonant sites visited, and thus
form less sparse fractals ~\cite{Yu}.

The same analysis was repeated at higher interaction strength $U=1.5$ and $U=3.0$, where the critical HF states appear away from the Fermi energy. The result is that  away from the Fermi energy $d_{2}$ is practically indistinguishable from the non-interacting case.

To assess the effect of the range of the interactions, we also computed $d_{2}$ at the Fermi level
when the Coulomb interaction was truncated as described in Sec.~2 (the critical interaction strength in this case was $U_{c}\approx 0.75$, a bit smaller than for the full-scale Coulomb interaction). With the Coulomb interaction restricted to 5 nearest neighbors in the Fock terms, the correlation dimension was $d_{2}\approx 1.39\pm 0.05$ both when the Coulomb interactions in the Hartree terms were restricted to 5 or to 20 nearest neighbors. This result shows that  Coulomb interactions of full range are essential to change the fractal dimension $d_{2}$  significantly. With truncated Coulomb interaction, the effective $d_{2}$ in a finite sample gradually decreases and approaches its non-interacting value.


\begin{figure}[h]
\center{\includegraphics[width=0.7\linewidth]{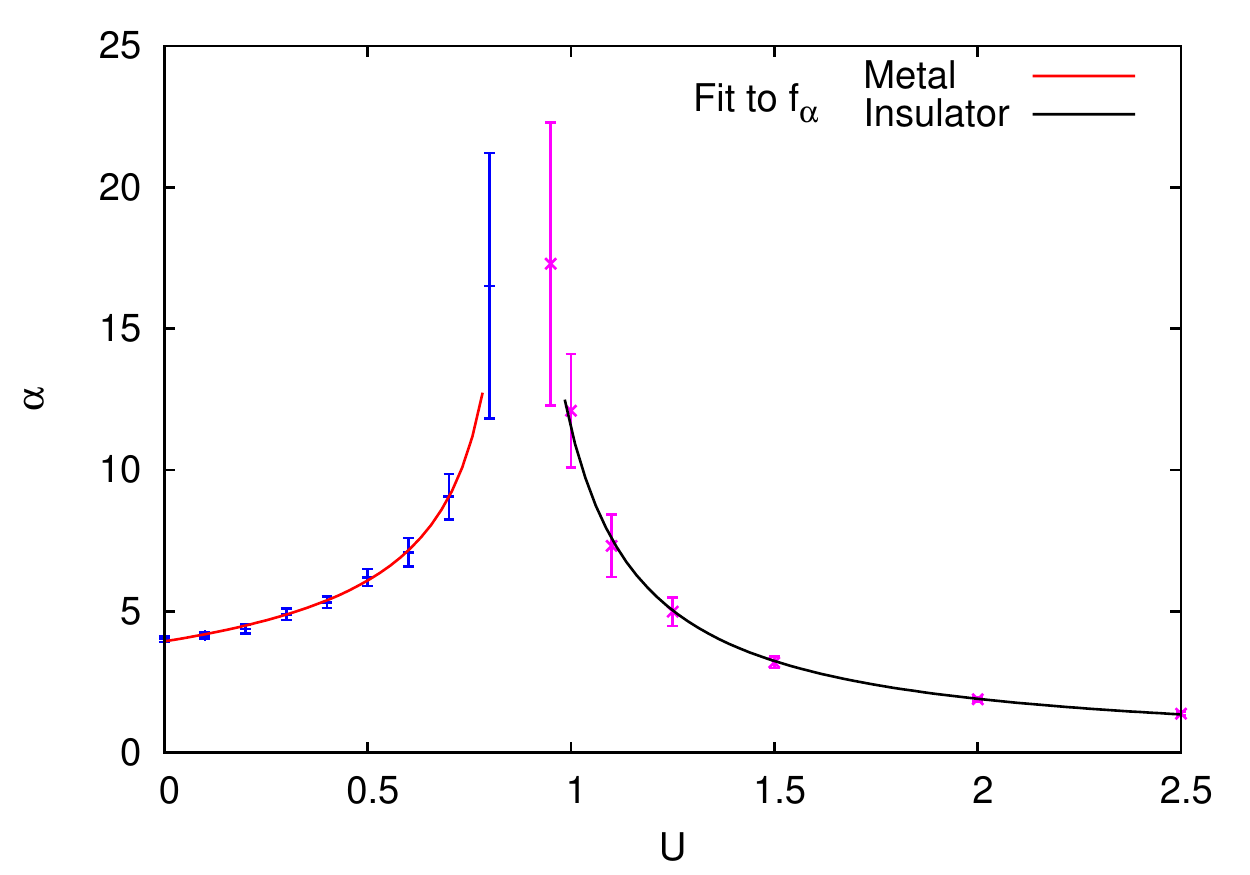}}
 \caption{(Color online) Evolution of the finite size
correlation length $\alpha$ with the interaction strength $U$ at $\varepsilon=\varepsilon_F$ ($W=14$ being fixed).
The raising and falling parts correspond to the metal and the insulator, respectively. In the interval $0.8<U<0.9$
the error-bars are too large to distinguish
between various scenarios discussed in the main text.
 }
\label{Fig:alpha}
\end{figure}

\section{ Metal-insulator transition.}
Fig.~\ref{Fig:alpha} shows the evolution
of the finite size-corrected correlation length $\alpha(U)$, as
obtained from the scaling collapse of Fig.~\ref{fig:K-collapse}. It exhibits strong
non-monotonicity, indicating a localization transition: For
$U<U_{<}\approx 0.79$, $\alpha(U)$ increases with increasing $U$ while for
$U>U_{>}\approx 0.89$, it decreases. The best fits to
critical power laws yield $\xi(U)=a\,|U-U_{<(>)}|^{-\nu_{M(I)}}$
with:
\be \label{nu-nu}\nu_{M}\approx 0.50\pm 0.05,\;\;\;\; \nu_{I}\approx
0.96\pm 0.05,\;\;\;\;\;{\rm full}-{\rm range}\;\;\;{\rm Coulomb}.\ee
The difference in the fit exponents is too big to be a mere result
of statistical errors, or of systematic errors related with the
fitting procedure. Indeed, as an independent check we
computed the critical exponents for a non-interacting system of
equal size, using the same method. We obtained \cite{fse}
  much closer exponents $\nu_{M}\approx
1.20\pm 0.05$, $\nu_{I}\approx 1.08\pm 0.08$. We thus believe that
the difference in the fit exponents~(\ref{nu-nu})
is a genuine interaction effect, which persists to fairly large scales. Possible interpretations of these findings are discussed further below.
In this context it is interesting
to note that the exponent $\nu_{M}\approx 0.5$ has been reported in
earlier experiments on ${\rm Si:P}$, which remained a puzzle for
theorists for a long time~\cite{Bel-Kirk}.

Eqs.~(\ref{nu-nu}) might reflect the degradation of the multifractal
pattern due to the interaction-induced mixing of non-interacting
wavefunctions, which we expect to be much stronger in the
delocalized than in the localized phase (where fewer non-interacting wavefunctions
involved
have a significant overlap). Another phenomenon that
undoubtedly influences the MI transition is the gradual breakdown of
screening in the metallic phase. The interactions, which are well
screened deep in the metal, must become long range somewhere on the
way to the insulator~\cite{LMNS}. This entails a crossover, or even a phase
transition, to a glassy phase~\cite{dobro,MuellerPankov}. We observe
a trace of the latter via the onset of non-uniqueness of the HF
solutions roughly at the same point as the MI
transition
but our resolution is not
sufficient to determine whether the two phenomena coincide. It also
remains an interesting open question whether screening breaks down
at the MI transition only, or already within the metal, as it
happens in mean field models with similar
ingredients~\cite{MuellerStrack}.

It is interesting to compare our results for the exponent $\nu_{I}$ with the $\epsilon$-expansion \cite{BurMirGor} obtained from the Finkel'stein's theory \cite{Fin-review} in the unitary ensemble. According to this theory:
\be
\nu^{{\rm inter}}=\frac{1}{\epsilon}-1.64,\;\;\;\;\nu^{{\rm non-inter}}=\frac{1}{2\epsilon}-\frac{3}{4}.
\ee
These expressions are meaningless at $\epsilon=1$, where they evaluate to negative $\nu$. One may think, however, that
for small $\epsilon$ they give a correct relationship between $\nu^{{\rm inter}}$ and $\nu^{{\rm non-inter}}$, as was the case for the fractal dimension $d_{2}$ in Eq.~(\ref{eps-d-2}).
 However, the relationship between $\nu^{{\rm inter}}$ and $\nu^{{\rm non-inter}}$ is ambiguous in the region of $\epsilon<1$ where both of them are still positive. Indeed, one can see that for very small $\epsilon$, one has $\nu^{\rm{inter}}>\nu^{{\rm non-inter}}$. However, as $\epsilon$ increases, $\nu^{{\rm non-inter}}$ catches up with
$\nu^{{\rm inter}}$, and at $\epsilon > 0.55$  we have $\nu^{\rm{inter}}<\nu^{{\rm non-inter}}$, as in our results for $d=3$. This may indicate that two competing mechanisms are at play,  whose relative importance depends on the dimensionality.

We also note that the exponent $\nu_{I}$ increases when the Coulomb interaction is truncated or when the mobility edge moves away from the Fermi level:
\be
\nu_{I}=1.31\pm0.1,\;\;(5^{{\rm th}}\; {\rm neighbors});\;\;\;\nu_{I}=1.36\pm 0.1,\;\;(\varepsilon_{m}-\varepsilon_{F}=1.7).
\ee
In contrast, the exponent $\nu_{M}$ is almost insensitive to truncation, but decreases with increasing interaction strength:
\be
\nu_{M}=0.40\pm0.03,\;\;\;(U=3,\varepsilon_{m}-\varepsilon_{F}=1.7).
\ee
The large error bars in the interval $U\in[0.8,0.9]$ do not allow us
to determine $\xi(U)$ by an accurate treatment of the finite-size
scaling $\alpha(U)=\xi(U)\,f_{\alpha}(\xi/L)$ \footnote{In our fitting procedure we used $f_{\alpha}(y)=[1+Cy^{1/\nu}]^{-\nu}$.}.  Two different
scenarii may be envisioned to reconcile the fit exponents
(\ref{nu-nu}) with standard theoretical considerations: (a) There is
a single localization transition close to $U_{c}=0.9$ with a
shoulder in the dependence of $\xi(U)$ on the metallic side, due to an additional phase transition
or a crossover in a different sector, such as the breakdown of
screening or the onset of glassiness. In that case $\nu_M$ would be
expected to approach $\nu_I$ sufficiently close to $U_c$ and on
large scales. (b) There are two separate transitions at
$U_{c1}\approx 0.79$ and $U_{c2}\approx 0.89$, with critical
wavefunctions in an entire finite interval $U\in[U_{c1},U_{c2}]$. In
this more exotic (and less probable) scenario, there would be no a priori reason for the
two exponents to coincide.

\section{ Finite-energy mobility edge in the insulator.}
A similar scaling analysis as above may be performed for $\varepsilon$
away from $\varepsilon_F$, which determines a critical line - the mobility edge
$\varepsilon_m(U)$. Of course, such a mobility edge is defined sharply only at the
mean field level of the HF equations, which neglect the finite life-time
of higher energy excitations due to inelastic processes involving either phonons or delocalized excitations of purely electronic origin
~\cite{AndersonFleishman}.
Nevertheless, the phase space
for decay processes at low energies is strongly suppressed, and due to the
pseudogap even more severely so than in an ordinary Fermi liquid.
This gives us confidence that the features of single particle HF
levels are representative of the fully interacting system. In
particular, the statement that $\varepsilon_m$  remains close to $\varepsilon_F$
is a result which we believe to be robust  beyond the Hartree-Fock
approximation. Hereby $\varepsilon_m$ should be interpreted as the (approximate) location where
the LDoS correlations become critical up to the relevant length scale set by inelastic decay processes.

\begin{figure}[h]
\center{\includegraphics[width=1.0\linewidth]{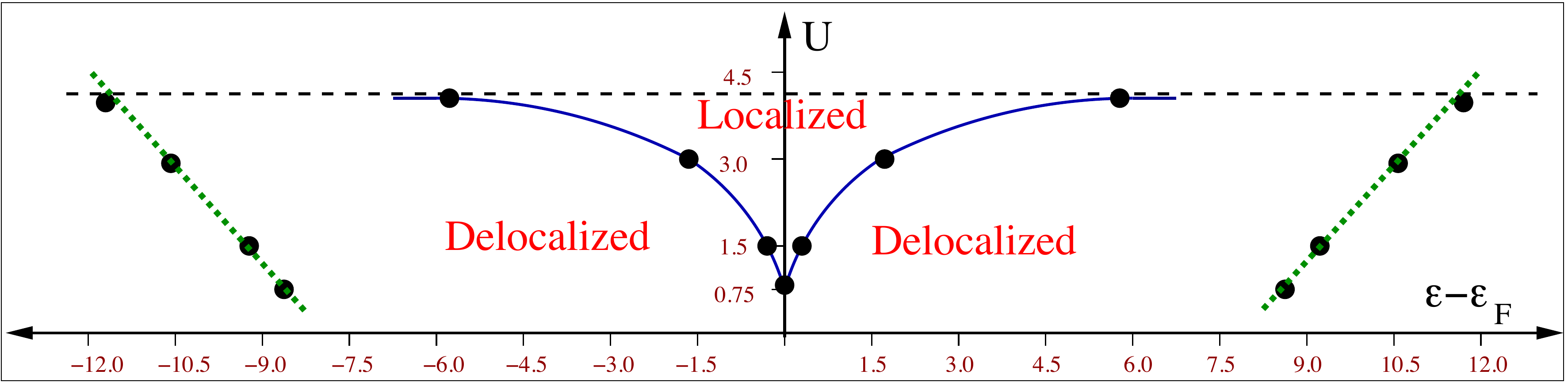}} \caption{(Color
online)  Phase diagram. In a wide range of the interaction strength
$U$ the mobility edge (solid blue line) stays close to
$\varepsilon_{F}$ as compared to the bandedges (green dashed line). This behavior is in a qualitative agreement with a conjecture \cite{BurMirGor-arX} that the mobility edge $U(\varepsilon-\varepsilon_{F})-U_{c}\propto (\varepsilon-\varepsilon_{F})^{\frac{1}{\eta\nu}}$ with $\nu\eta\approx 2$.}
\label{fig:sketch}
\end{figure}

Our result is shown in Fig.~\ref{fig:sketch} together
 with the bandedge, defined as the energy where $\rho(\varepsilon)$ drops to half of its maximal value.
Fig.~\ref{fig:sketch} demonstrates that the mobility edge is indeed
trapped in a narrow range around $\varepsilon_F$. This holds for values of $U$ nearly all
the way up to $U_*\approx 4$ where the last states around the
maximum of the DoS localize (at $W=14$). This confirms the
expectations of Ref.~\cite{Yazdani} that in a relatively broad
region of the parameters $W$ and $U$, states near $\varepsilon_F$ are
almost critical. Note also that $U_*$ is almost 5 times larger than
$U_c\approx 0.85$ where the MI transition occurs. In fact $U>
U_*$ brings the system already  very close to a Mott-type transition
where charge-density wave order sets in. It remains an interesting
question to study how such charge ordering effects and glassiness
(i.e., the multiplicity of HF solutions) affect the localization as
the interaction strength increases.

\section{Multiplicity of HF solutions: glassiness and charge ordering.}
\label{s:glass}
Finally we briefly address the issue of multiple solutions of the
Hartree-Fock equations and its relation to the onset of glassy
behavior as the interaction $U$ increases.

\begin{figure}[h]
\center{\includegraphics[width=0.5\linewidth]{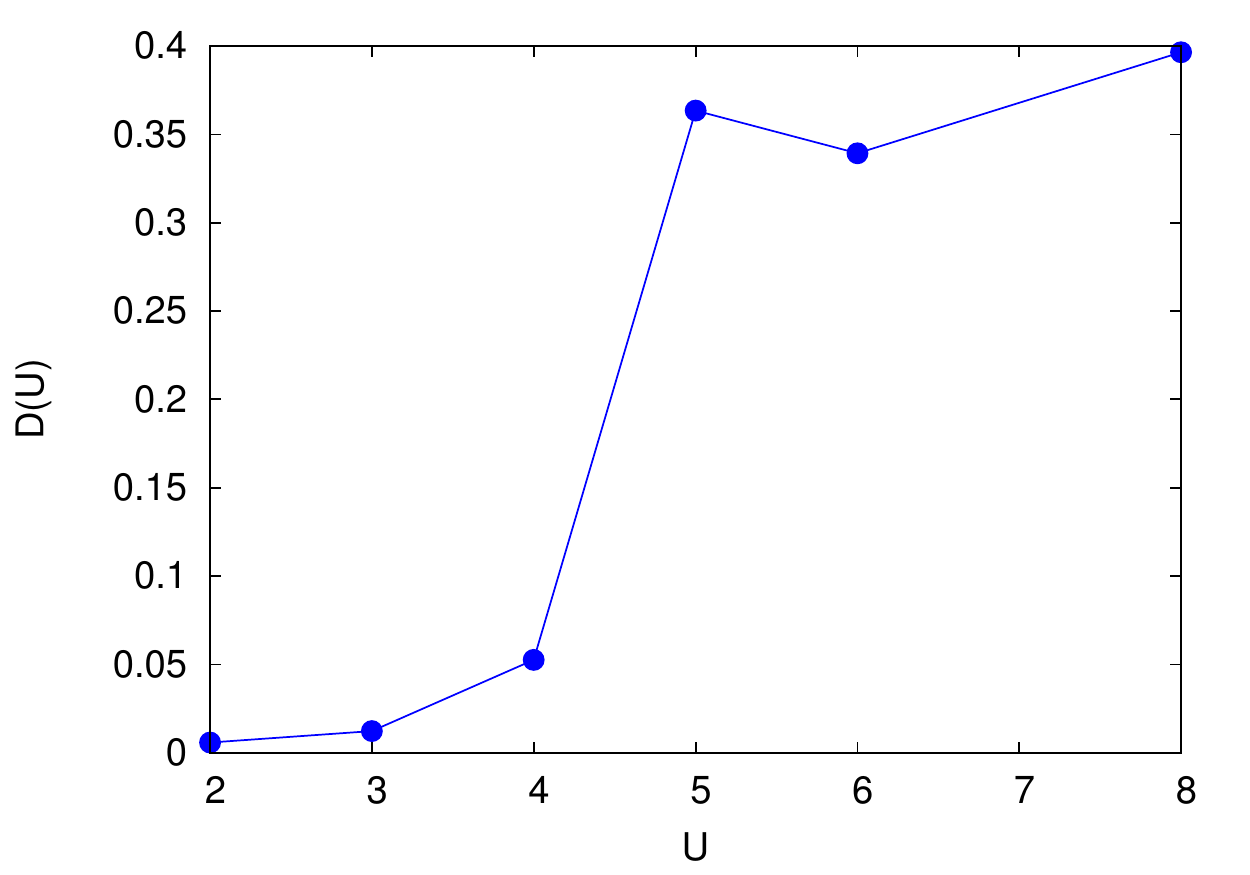}}
\caption{Average variance of on-site occupation numbers $n({\bf r})$
as a function of interaction strength.} \label{fig:sol}
\end{figure}

Our iterative procedure to solve the HF equations begins with an
input configuration of occupation numbers $n_{\rm in}({\bf r})$ on
each of the $N$ lattice sites. At sufficiently large $U$ the
converged output HF solution $n_{\rm out}({\bf r})$ is generally
different for runs with different inputs. To quantify this
difference statistically we studied the quantity:
\be\label{form} D(U)=\frac{1}{N_{\rm sol}}\sum_{m}\frac{1}{N}\sum_{{\bf
r}}|n_{\rm out}^{(m)}({\bf r})-n_{\rm out}^{(0)}({\bf r})|^{2},
\ee
where the superscript $m$ labels the set of $N_{\rm sol}$ different solutions which were obtained from
initial density patterns $n_{\rm in}^{(m)}({\bf r})$, while $0$ denotes a
reference solution. In the simplest test we have chosen $N_{\rm sol}=10$ solutions, out of
which 8 were obtained from random inputs and 2 had a checkerboard order as
input. The results for $D(U)$ are presented in Fig.~\ref{fig:sol}. One can see that for $U<4$ the average deviation of the solutions from
the reference solution is small. It is thus reasonable to assume that
physical properties evaluated on the various solutions are
statistically very similar. However, starting from $U\approx 4$ the
function $D(U)$ sharply increases.
\begin{figure}[h]
\center{\includegraphics[width=1.0\linewidth]{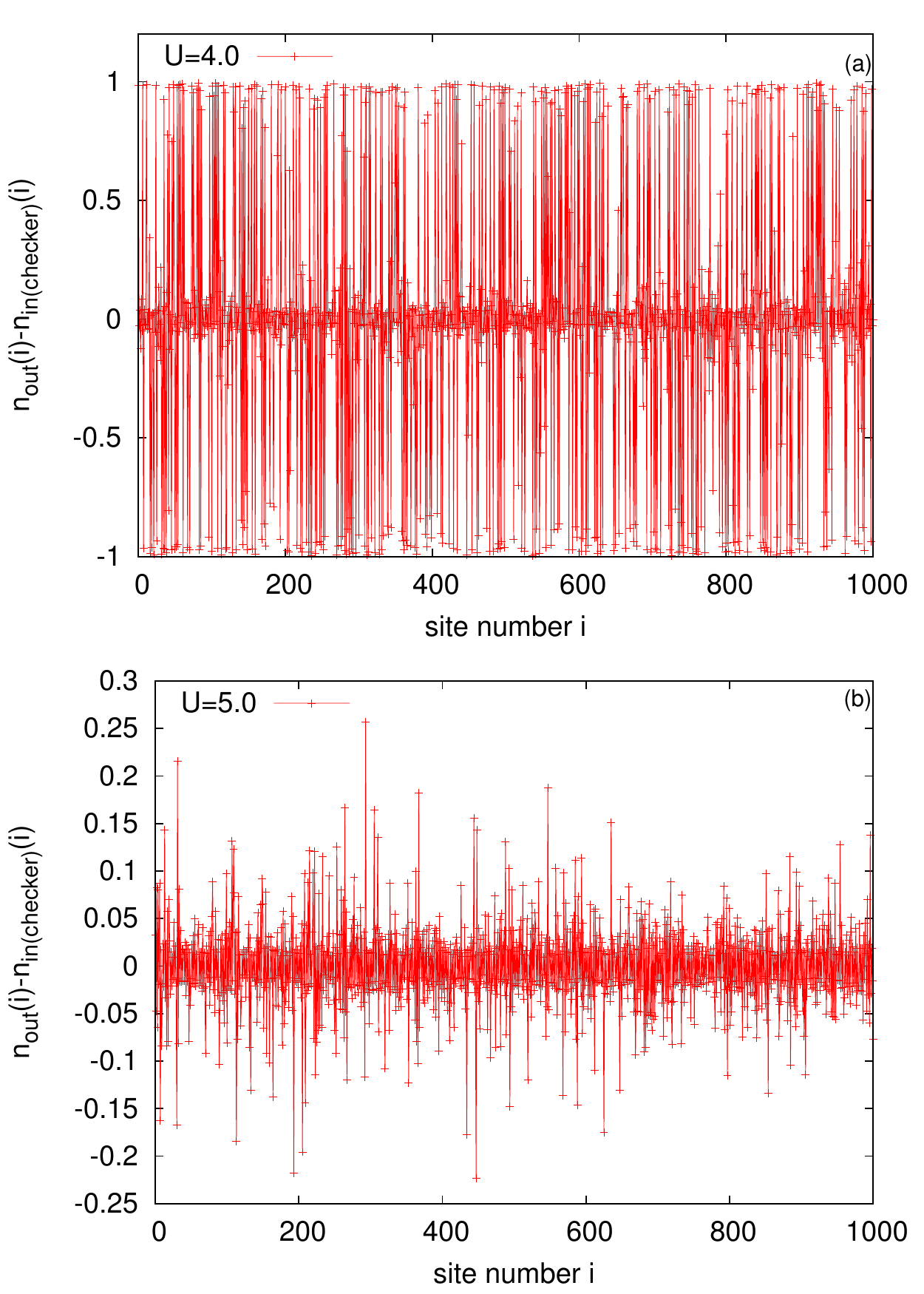}}
\caption{Difference between the output on-site occupation number and
the checkerboard input occupation number: (a) $U=4$ output is random
and uncorrelated with the input, (b) $U=5$ output has almost the
same checkerboard structure as the input.  } \label{fig:checker}
\end{figure}

In order to check whether this increase is due to a significant
variation between {\it random} HF solutions or whether this increase
in $D(U)$ is due to stabilization of a checkerboard density pattern
in the HF solution, we consider the solution obtained from
initializing with a checkerboard input and plot the difference
between the solution and the corresponding input pattern.

The result for $U=4$ shows (see Fig.~\ref{fig:checker}(a)) that the
difference has a clear checkerboard structure which implies that the
output was random. However, as $U$ increases to $U=5$ the difference
reduces significantly (see Fig.~\ref{fig:checker}(b)), which signals
the tendency to retain the checkerboard order in the solution.
Comparing also with free energies of random solutions, we concluded
that the transition to a checkerboard structure (charge density
wave) occurs somewhere in the range $4< U <5$.

A more precise identification of the onset of multiplicity of
solutions shows that it starts at much smaller values $U\approx
0.7$, which roughly coincides with the $U_{c}$ at which localization
at the Fermi energy occurs \footnote{these preliminary results are obtained with the Coulomb interaction truncated to $5^{{\rm th}}$ neighbors in the Fock terms and to $20^{{\rm th}}$ neighbors in the Hartree terms}. In order to show this we generated 10
different HF solutions at $U=1.0$ and characterized them globally by
the total energy $E$ per site. The fact that the iterative HF
procedure at $U=1.0$ converges to different values of $E$ is a
manifestation of the existence of multiple local minima. Physically,
one may expect that this will reflect in the onset of glassy
behavior associated with the slow dynamics or relaxation between
the minima that correspond to the various HF solutions.

\begin{figure}[h]
\includegraphics[width=1.0\linewidth]{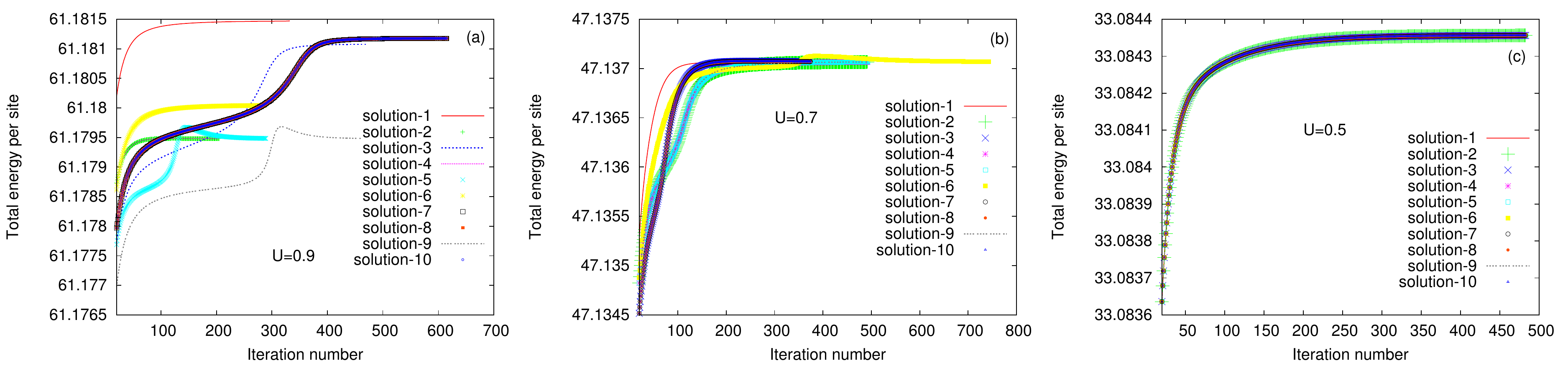}
\caption{Convergence of the HF procedure for the total energy per
site: (a) $U=0.9$, the total energy converges to different output
values, depending on the initial configuration of on-site occupation
numbers; (b) $U=0.7$, different initial configurations of occupation
numbers obtained in the previous step converge to the same value of
the total energy, and thus to the same HF solution; (c) $U=0.6$ no
further divergence in the total energy is observed any more.}
\label{fig:onset}
\end{figure}

We then used the above solutions as inputs at slightly decreased
interaction strength $U=0.9$, the resulting solutions still
being of different total energy (see Fig.~\ref{fig:onset}(a)). Upon
decreasing the interaction in steps of $0.1$, and using the
solutions of the previous step as initial condition, we found that
at  $U\approx 0.7$ the total energies, after a large number of
iterations, coincided (Fig.~\ref{fig:onset}(b)). That value of $U$ can thus
be interpreted, at this HF mean field level, as the border of a
glassy regime.  Upon further decrease of $U$ the solutions did not
diverge anymore, implying the existence of a unique HF solution
(Fig.~\ref{fig:onset}(c)).

The fact that in the insulating regime the HF equations develop a
number of solutions which grows exponentially with the volume is to
be expected, as this is well-known to be the case in the classical
limit of vanishing hopping, $t=0$. One may wonder, whether and how
 key features like the Efros-Shklovskii Coulomb gap are present in {\em typical}
solutions to the HF equations, or whether they occur only in the lowest energy solutions, which are very difficult to find. We argue here, that all typical
solutions are expected to exhibit a parabolic Coulomb gap, as we
indeed observed numerically.

To understand how such a Coulomb gap comes about, consider the HF Fock equations
in the limit of vanishing hopping, $t=0$, where all HF orbitals are
completely localized.
 A HF solution consists in an assignment of
occupation numbers $n_i \in \{0,1\}$ to the sites $i$, according to
whether the local potential,

\bea E_i = \epsilon_i +\sum_j  \frac{n_j}{r_{ij}} \eea
is above ($\to n_i=0$) or below ($\to n_i =1$) the chemical
potential $\mu$ ($\mu \approx 0$ is always adjusted to assure half
filling). In the classical limit,  the HF procedure consists in
updating the occupation numbers until convergence to a stable point
is reached. The final HF solution is a minimum of the HF energy with
respect to the change of any of the $n_i$, if the HF energy is
written in grand-canonical form, including the term $-\mu\sum_i
n_i$. That is, there is stability with respect to single particle
addition or removal.

As long as there is no suppression in the low energy distribution of
the local potentials $E_i$ there are lots of rearrangements in each
update. Those die out only once at least a parabolic pseudo-gap
develops in the distribution of the $E_i$'s, such that the probability of a change of occupation triggering other rearrangements becomes small. Note that the $E_i$'s
are just the classical limit of the HF eigen-energies. Thus the
convergence of the HF procedure guarantees essentially the presence
of a Coulomb gap in the LDOS. This happens, even though we do not
impose explicitly the stability of HF solutions with respect to
single particle moves, i.e., swaps between configurations $(n_i,n_j)
=(0,1)$ and $(1,0)$. The latter are the elementary moves considered
in standard arguments for the Efros-Shklovskii Coulomb gap in classical Coulomb
glasses, but the above reasoning shows that one does not need to impose that extra
stability constraint to obtain a well-developed Coulomb
gap.~\footnote{Similar observations were made by A. Amir,
M.~Palassini, B.~Shklovskii and B. Skinner, private discussion}.

When, on top of the HF equations, stability with respect to
particle-hole excitations and more complex rearrangements is
imposed, the Coulomb gap is hardening a bit, but no essential new
features appear~\cite{Moebius}. For this reason we contented ourselves
with an analysis of {\em typical} solutions of the HF equations, without
further minimizing the HF energy among the
exponentially many solutions. We expect that the localization
properties, multifractality etc. evolve only very weakly as one
biases the considered HF solutions towards lower-lying and more
stable solutions. Indeed, our scalings work well when evaluating them in
typical solutions on the insulating side, and key physical
observables behave as we expected, even in the limit $t=0$.

\section{ Conclusion.} We have studied numerically the localization
transition in a 3D Anderson model of spinless fermions, with Coulomb
interactions treated within the HF approximation. The metal-insulator transition was
identified via the localization at the Fermi level, determined from
a detailed study of the auto-correlation function of the HF
eigenfunctions. Our main results are: {\em (i)} Multifractal power
law scalings in the local DoS
survive the presence of interactions, and extend up to a (large) correlation length $\xi(U,\varepsilon)$.
{\em (ii)} A critical Coulomb gap in the weakly insulating phase pins the mobility edge close
 to $\varepsilon_F$ for a wide range of parameters, while most higher energy excitations are still
 delocalized. At disorder strength $W=14$ (moderately close, but not fine-tuned, to the non-interacting
critical disorder $W=16.5$) the critical $U_{c}(\varepsilon_F)$ for
the metal-insulator transition is $\sim 5$ times smaller than the $U_*$ required
for  localization of the entire HF spectrum. This is in qualitative agreement with
the experimental observations of Ref.~\cite{Yazdani}.
 {\em (iii)} A scaling analysis of the DoS reveals a critical Coulomb anomaly $\rho(\omega)\sim\omega^{0.6\pm0.15}$, and scaling laws as anticipated in Refs.~\cite{McMillan,LMNS}.
 {\em (iv)} The apparent correlation length exponents display a significant asymmetry between the metallic and
 insulating sides, similar to tendencies reported in experiments. We conjecture that they arise from
 crossover phenomena in the metallic phase related with the breakdown of screening or the onset of glassy
 metastability seen in HF.
 Those deserve further future studies.

\subsection*{Acknowledgments}

 We thank I. Girotto for help with parallel
programming and  A. Yazdani,  M. Feigel'man and B. I.  Shklovskii for stimulating discussions. 
MA is grateful to S. A. Jafari and F. Shahbazi for
useful comments and interest in this work and to the CM\&SP section of
ICTP for hospitality.


\section*{References}
\begin{thebibliography}{99}
\bibitem{Yazdani} A. Richardella et al., Science {\bf 327}, 665
(2010).
\bibitem{50Anderson} \emph{50 Years of Anderson Localization}, Edited by E. Abrahams,
World Scientific Publishing Co Ltd,  Singapore, 2010.
\bibitem{ES} A. L. Efros and B. I. Shklovskii,
J. Phys. C: Solid State Phys. {\bf 8}, L49 (1975).
\bibitem{ES-Springer} A. L. Efros and B. I. Shklovskii, \emph{Electronic properties of doped
semiconductors}, Springer series in solid-state sciences, Springer,
1984.
\bibitem{AA-Elsevier-book} \emph{Electron-electron interactions in disordered systems},
Edited by B. L. Altshuer, A. G. Aronov, A. L. Efros and M. Pollak,
Elsevier, North Holland, 1985.
\bibitem{BurMirGor} I. S. Burmistrov, I. V. Gorny, A. D. Mirlin, Phys. Rev. Lett. {\bf 111}, 066601 (2013).
\bibitem{Fin-review} A. M. Finkel'stein, Int. J. Mod. Phys. B
{\bf 24}, 1855 (2010).
\bibitem{Bel-Kirk}  D. Belitz, T. R. Kirkpatrick,
Rev. Mod. Phys.  {\bf 66},  261 (1994).
\bibitem{LevShyt}
L. S. Levitov, A. V. Shytov, Pis'ma Zh. Eksp. Teor. Fiz. {\bf 66},
200 (1997) [JETP Lett., {\bf 66}, 214 (1997).
\bibitem{KamenevAndreev} A. Kamenev and A. Andreev, Phys. Rev. B  {\bf 60},
2218 (1999).
\bibitem{BES84} S. D. Baranovskii, B. I. Shklovskii, and A. L. Efros, Zh. Eksp. Teor. Fiz. {\bf 87}, 1793 (1984). [Sov. Phys. JETP {\bf 6}, 1031 (1984).]
\bibitem{LMNS} M. Lee, J. G. Massey, V. L. Nguyen, B. I. Shklovskii,
Phys. Rev. B {\bf 60}, 1582 (1999).
\bibitem{cond-mat} M. Amini, V. E. Kravtsov and M. M\"uller, arXiv:1305.0242v1.
\bibitem{Vojta} F. Epperlein, M. Schreiber and T. Vojta, Phys. Rev.
B {\bf 56}, 5890 (1997).
\bibitem{Mirlin-rep}  F. Evers and A. D. Mirlin, Rev. Mod. Phys. {\bf 80},
1355 (2008).
\bibitem{KrCue} E. Cuevas and V. E. Kravtsov, Phys. Rev. B {\bf 76},
235119 (2007).
\bibitem{IFK} M. V. Feigel'man, L.B. Ioffe,V. E. Kravtsov, E. A. Yuzbashyan, Phys. Rev. Lett. {\bf 98} 027001
(2007).
\bibitem{IFKCue} M. V. Feigel'man et al., Ann. Phys. {\bf 365} 1368
(2010).
\bibitem{Bur} I. S. Burmistrov, I. V. Gornyi, A. D. Mirlin, Phys. Rev. Lett. {\bf 108}, 017002
(2012).
\bibitem{KrProc} V. E. Kravtsov, J. Phys.: Conf. Ser. {\bf 376}, 012003
(2012).
\bibitem{Lee} P. A. Lee, Phys. Rev. B 26, 5882 (1982)
\bibitem{16-5} T. Ohtsuki, K. Sleving and T. Kavarabayashi, Ann.
Phys. (Leipzig), {\bf 8}, 5, (1999); K. Slevin and T. Ohtsuki, Phys.
Rev. Lett., {\bf 82}, 382 (1999).
\bibitem{McMillan} W. L. McMillan, Phys. Rev. B {\bf 24}, 2739 (1981).
 \bibitem{LeePRL} M. Lee, Phys. Rev. Lett., {\bf 93}, 256401 (2004).
\bibitem{fse} The mobility edge  $W_{c}=19\pm 1.0$  and the exponent $\nu=1.15\pm
0.1$ found like this
differ
from those ($W_{c}=16.54\pm 0.02$, $\nu=1.57\pm 0.02$)
of earlier studies of conductance \cite{16-5}.
We attribute this discrepancy to finite-size corrections.
Those strongly
affect the exponent $\nu$ even in much bigger samples~\cite{16-5}.
\bibitem{MuellerPankov} M.~M\"uller and S. Pankov, Phys. Rev. B {\bf 75}, 144201 (2007).
\bibitem{dobro} V. Dobrosavljev\'ic, D. Tanaskovic, and A. A. Pastor, Phys. Rev. Lett. {\bf 90}, 016402 (2003).
\bibitem{MuellerStrack} M.~M\"uller, P. Strack and S. Sachdev, Phys. Rev. A {\bf 86}, 023604 (2012).
\bibitem{BurMirGor-arX}  I. S. Burmistrov, I. V. Gornyi, A. D. Mirlin, arXiv: 1307.5811.
\bibitem{AndersonFleishman} L. Fleishman, P. W. Anderson, Phys. Rev. B {\bf 21}, 2366 (1980).
 \bibitem{Moebius} A. M\"obius, M. Richter, and B. Drittler, Phys. Rev. B
{\bf 45}, 11 568 (1992).

\bibitem{Yu} X. Yu, "Superfluidity and localization in Bosonic glasses", SISSA,
Scuola Internazionale Superiore di Studi Avanzati, 2012.
URI: http://hdl.handle.net/1963/6327.
V. Bapst, X. Yu , and M. M\"uller, {\em in preparation}.


\end{thebibliography}
\end{document}